\documentclass[journal]{IEEEtran}
\usepackage{amsmath,amssymb,amsthm}
\usepackage{graphicx}
\usepackage{booktabs}
\usepackage{cite}
\usepackage{xcolor}
\usepackage[hidelinks]{hyperref}
\newtheorem{lemma}{Lemma}
\newcommand{\E}{\mathbb{E}}
\newcommand{\bH}{\mathbf{H}}
\newcommand{\bR}{\mathbf{R}}
\newcommand{\bx}{\mathbf{x}}

\begin{document}
\title{Timestep Selection in Diffusion-Based OTFS Channel Estimation}
\author{Yi Luo, Dianxin Zheng, Ruixuan Mi, Xinqi Fu, and Tianhao Guo*
\thanks{Corresponding author: Tianhao Guo (e-mail: your-email@example.com).}
\thanks{Code and trained networks: \url{https://github.com/tellsgo/otfs-timestep-selection}, archived at [Zenodo DOI].}
}
\maketitle

\begin{abstract}
Diffusion-based channel estimators are run with a few deterministic sampling steps, and their step count is treated as a knob that trades accuracy against latency. We show that for the mean squared error this trade-off mainly comes from the default leading timestep spacing, whose first timestep indicates an intermediate noise level although the initial state is pure noise. In a multiple-input multiple-output orthogonal time frequency space receiver with ray-traced paths and per-path Doppler shifts, moving only the first timestep of the default spacing to the pure-noise end recovers the gain of trailing spacing, completely for our anchored residual estimator and mostly for a published diffusion estimator. With trailing spacing one to three steps give the most accurate single trajectory of our estimator at every signal-to-noise ratio we test, and longer trajectories raise the error, by 0.34~dB over one step at a signal-to-noise ratio of 20~dB after 50 steps. This is the direction predicted by an idealized analysis in which a long deterministic trajectory turns the posterior mean into a posterior sample. With three trailing steps, the published estimator reaches 0.8 to 1.8~dB lower normalized mean squared error than with its original five leading steps, on its own trained network and test set. A start rule that places the first timestep from the error power of the linear anchor of the estimator, with one threshold per network, recovers trailing spacing for our estimator and matches start timesteps tuned separately for every signal-to-noise ratio on a noise-prediction network. A regression network with the same architecture and data reaches 0.23 to 1.36~dB lower normalized mean squared error than the best diffusion configuration we evaluate. The step count of a diffusion channel estimator should therefore be judged only after its first timestep is fixed.
\end{abstract}


\section{Introduction}
\IEEEPARstart{O}{rthogonal} time frequency space (OTFS) modulation \cite{hadani2017otfs,wei2021otfs} is designed for high-mobility links, and its receiver needs an estimate of the doubly dispersive channel in the delay-Doppler (DD) domain. Score-based and denoising diffusion models are increasingly used for this kind of channel estimation. They learn the channel distribution from ray-traced or measured data and estimate the channel by running a reverse diffusion process conditioned on pilot observations \cite{arvinte2023mimo,fesl2024diffusion}. The low-complexity estimator of \cite{fesl2024diffusion} starts this process at the timestep whose noise level matches the signal-to-noise ratio (SNR) of the pilot observation, initializes it with the normalized observation and skips the noisier steps. For OTFS, the conditional residual diffusion model (CRDM) of \cite{zheng2026crdm} refines an embedded-pilot least-squares estimate in the DD domain with deterministic steps of the denoising diffusion implicit model (DDIM) \cite{song2021ddim}. CRDM selects five steps as its default and keeps them at every speed, and its speed-averaged accuracy-latency curve reports a loss of about 1.3~dB for three steps. With its published network and test set we measure this three-step loss as 0.64 to 0.86~dB at SNRs from 0 to 10~dB under the same leading spacing (Section~\ref{sec:abl}). In this line of work more steps buy accuracy at the price of latency.

This paper examines that premise for the mean squared error (MSE) by which channel estimators are judged. Leading spacing, the default of common DDIM implementations, starts the reverse process at an intermediate timestep although the initial state is pure Gaussian noise, a flaw shown for image generation in \cite{lin2024flawed}, where it biases the brightness of the generated samples. For estimation this flaw inflates the number of steps that appear necessary. We study a newly trained anchored residual diffusion estimator, which refines a linear minimum mean squared error (LMMSE) anchor on channels with per-path Doppler shifts, and we evaluate the published CRDM network on its own test set with its trained weights unchanged to compare gains. Two mechanisms set the step count. Under leading spacing the noise level indexed by the first timestep does not match the network input, which costs accuracy at small step counts, so more steps help. Once the first timestep sits at the pure-noise end, this mismatch disappears, and a longer trajectory shifts the estimate from the posterior mean toward a posterior sample, so more steps no longer improve our estimator and eventually raise its error. The first timestep is the primary contributor to the loss at small step counts, and the published step count can therefore be re-examined by changing only the sampler, without retraining. The contributions of this paper are as follows.
\begin{itemize}
\item Moving only the first leading timestep to the pure-noise end shows that this timestep mainly causes the loss of the default leading spacing at small step counts. With $T=1000$ timesteps and $t=0$ the clean end, leading spacing starts the reverse process at $t=800$ for five steps and at $t=0$ for one step, while the initial state is pure Gaussian noise. With the trailing spacing of \cite{lin2024flawed}, which starts at $t=999$, one to three steps give the most accurate single trajectory at every SNR considered.
\item Additional steps raise the error. For an idealized sampler the one-step estimate is the posterior mean, whereas a long deterministic trajectory yields a posterior sample, which adds the posterior variance to the error of the mean and so doubles the minimum mean squared error (MMSE). The measured increase over one step grows with the SNR and reaches 0.34~dB at an SNR of 20~dB after 50 steps, the largest step count considered.
\item An anchor-error-matched start chooses the first timestep from the error power of the LMMSE anchor at the operating SNR, which reflects the pilot SNR, the window truncation and the fractional-Doppler leakage, with one threshold per network chosen on validation frames. For our $v$-prediction network it selects the trailing start itself. For a noise-prediction network, on which every tested start at the pure-noise end gives a normalized mean squared error (NMSE) of $+15$ to $+17$~dB, it matches a start tuned per SNR with one parameter. At SNRs not used for calibration it comes within 0.05~dB of the oracle start chosen on the test frames, although that network is less accurate than its anchor.
\item Under leading spacing the best step count is 50 up to 15~dB and 10 at 20~dB, where 5 to 50 steps lie within 0.06~dB. Under trailing spacing one to three steps are best, and at a comparable number of network evaluations eight independent one-step estimates attain 0.08 to 0.47~dB lower NMSE than one ten-step trajectory at SNRs from 5 to 20~dB.
\item The published CRDM network gains up to 1.99~dB from trailing spacing with five steps, moving its first timestep alone recovers most of this gain, and its best trailing step count remains five or ten. A regression network of the same architecture, at the cost of one diffusion step, attains 0.23 to 1.36~dB lower NMSE than the best diffusion configuration we evaluate on the test cities and 0.22 to 1.53~dB lower NMSE on the unseen city, and we also compare the estimators in bit error rate (BER). We recommend that diffusion channel estimators be reported with their first timestep or start rule and with such a regression baseline.
\end{itemize}
The rest of this paper is organized as follows. Section~\ref{sec:sys} gives the system model and the estimator. Section~\ref{sec:spc} compares leading and trailing spacing, contrasts the posterior mean with a posterior sample, discusses the step count and introduces the anchor-error-matched start. Section~\ref{sec:res} reports numerical results, including the BER and the published CRDM network, and Section~\ref{sec:con} concludes the paper.

\emph{Notation:} $\E[\cdot]$ denotes expectation, $\|\cdot\|$ the Euclidean norm of an array stacked into a vector, $(\cdot)^*$ complex conjugation, $(x)_N$ the residue of $x$ modulo $N$, $\mathbf{I}$ the identity matrix and $\mathcal{N}(\mathbf{0},\mathbf{I})$ the real standard Gaussian distribution.

\section{System Model and Anchored Residual Diffusion Estimator}
\label{sec:sys}
\subsection{MIMO-OTFS Link}
A base station with $N_t=4$ transmit antennas serves a user with $N_r=4$ receive antennas, both half-wavelength uniform linear arrays, at a carrier frequency of 3.5~GHz. Each OTFS frame occupies $M=128$ delay bins and $N=64$ Doppler bins with subcarrier spacing $\Delta f=15$~kHz, so the delay resolution is $\Delta\tau=1/(M\Delta f)=521$~ns and the Doppler resolution is $\Delta\nu=\Delta f/N=234$~Hz. Table~\ref{tab:par} lists the simulation parameters. With ideal pulses the received DD signal at antenna $r$ is
\begin{equation}
y_r[k,l]=\sum_{u=1}^{N_t}\sum_{k',l'} h_{ru}[k',l']\,d_u[(k-k')_N,(l-l')_M]+w_r[k,l],
\label{eq:io}
\end{equation}
where $k$ and $l$ index Doppler and delay, $d_u$ is the DD symbol grid of transmit antenna $u$, and $w_r[k,l]$ is white Gaussian noise of variance $\sigma^2$. The SNR is $1/\sigma^2$ for unit-energy data symbols and channels of unit mean link energy. The DD response of link $(r,u)$ is a superposition of ray-traced paths,
\begin{multline}
h_{ru}[k,l]=\sum_{p} g_p\,a_{r}(\boldsymbol{\theta}^{\mathrm{A}}_p)\,a^*_{u}(\boldsymbol{\theta}^{\mathrm{D}}_p)\\
\times D_N(k-\nu_p/\Delta\nu)\,D_M(l-\tau_p/\Delta\tau),
\label{eq:channel}
\end{multline}
with path gain $g_p$, array responses $a_r,a_u$ at the angles of arrival (A) and departure (D), delay $\tau_p$, Doppler shift $\nu_p=(V/\lambda)\sin\theta^{\mathrm{A}}_p\cos(\phi^{\mathrm{A}}_p-\psi)$ for zenith and azimuth angles of arrival $\theta^{\mathrm{A}}_p,\phi^{\mathrm{A}}_p$, wavelength $\lambda$, user speed $V$ and heading $\psi$, and Dirichlet kernels $D_N(x)=\frac1N\sum_{n=0}^{N-1}e^{j2\pi nx/N}$ that spread fractional delays and Doppler shifts over the grid. Path parameters come from the DeepMIMO v3 city scenarios \cite{alkhateeb2019deepmimo}. Every path carries its own Doppler shift, so the channel is doubly dispersive rather than a delay profile shifted by a common Doppler shift as in the test set of \cite{zheng2026crdm}.

\begin{table}[t]
\centering
\caption{Simulation Parameters}
\label{tab:par}
\footnotesize
\setlength{\tabcolsep}{3pt}
\begin{tabular}{ll}
\toprule
Parameter & Value \\
\midrule
Carrier frequency & 3.5~GHz \\
Antennas $N_t$, $N_r$ & 4, 4 \\
Arrays & half-wavelength linear \\
Delay bins $M$, Doppler bins $N$ & 128, 64 \\
Subcarrier spacing $\Delta f$ & 15~kHz \\
Delay and Doppler resolution & 521~ns, 234~Hz \\
User speed $V$ & 30 to 500~km/h \\
Channel & DeepMIMO v3, eight cities, New York \\
Data symbols & QPSK \\
Pilot SNR & 20~dB above the SNR \\
Pilot bins & Doppler 32, delay 40, 44, 48, 52 \\
Window per pilot & 15 Doppler $\times$ 4 delay bins \\
Frames for the LMMSE anchor & 16{,}384 per SNR \\
SNR grid of the anchor & $-5$ to 25~dB in 5~dB steps \\
Timesteps $T$, noise schedule & 1000, cosine \\
U-Net size & 8.26~M real parameters \\
Training & 60{,}000 iterations, batch of 64 frames \\
Noise-prediction network & 30{,}000 iterations, linear schedule \\
Test frames & 512 or 1024 (Sec.~\ref{sec:setup}) \\
\bottomrule
\end{tabular}
\end{table}

Each transmit antenna sends one embedded pilot of energy $E_p$ at its own delay offset inside a shared guard region \cite{raviteja2019embedded,shen2019massive}, so the pilot SNR exceeds the SNR defined above by 20~dB. The receiver reads a $(2k_{\max}+1)\times(l_{\max}+1)$ window around each pilot, with $k_{\max}=7$ and $l_{\max}=3$, that is, 15 Doppler and four delay bins. The 15 Doppler bins cover the nominal Doppler span at 500~km/h. The windows collect noise, leakage of neighboring pilots and data leakage caused by fractional Doppler shifts, and the channel leakage outside them is extrapolated by the anchor of Section~\ref{sec:anchor}.

\subsection{Anchored Conditional Residual Diffusion}
\label{sec:anchor}
Let $\mathbf{y}$ stack all windows of all antennas and let $\hat{\bH}_{\mathrm{A}}=\mathbf{W}\mathbf{y}$ be the LMMSE estimate of the whole DD channel tensor $\bH$, which we call the anchor, with $\mathbf{W}$ built from uncentered second moments of 16{,}384 training frames per SNR (Section~\ref{sec:setup}), which include all interference terms. A diffusion model learns the distribution of the normalized residual $\bR=(\bH-\hat{\bH}_{\mathrm{A}})/s$ on $T=1000$ timesteps, where $s$ is the root mean square of the residual over training frames, fixed before training, and $t=0$ is the clean end. The cosine noise schedule \cite{nichol2021improved} sets the cumulative signal power coefficient $\bar\alpha_t$ of the forward state $\bx_t=\sqrt{\bar\alpha_t}\bR+\sqrt{1-\bar\alpha_t}\boldsymbol{\epsilon}$ with standard Gaussian noise $\boldsymbol{\epsilon}$, for $t=0,\dots,T-1$. A U-Net $f_\theta$ with 8.26~M real parameters receives $\hat{\bH}_{\mathrm{A}}/s+\bx_t$, the timestep and the SNR, and predicts $\mathbf{v}_t=\sqrt{\bar\alpha_t}\boldsymbol{\epsilon}-\sqrt{1-\bar\alpha_t}\bR$ \cite{salimans2022progressive}, which we call $v$-prediction. Writing its output as $\hat{\mathbf{v}}=f_\theta(\bx_{t_i},t_i)$, with the anchor and SNR inputs suppressed, the implied predictions of the residual and of the noise are $\hat{\bx}_0=\sqrt{\bar\alpha_{t_i}}\bx_{t_i}-\sqrt{1-\bar\alpha_{t_i}}\,\hat{\mathbf{v}}$ and $\hat{\boldsymbol{\epsilon}}=\sqrt{1-\bar\alpha_{t_i}}\bx_{t_i}+\sqrt{\bar\alpha_{t_i}}\,\hat{\mathbf{v}}$, and each of the $S$ deterministic DDIM steps updates
\begin{equation}
\bx_{t_{i+1}}=\sqrt{\bar\alpha_{t_{i+1}}}\,\hat{\bx}_0+\sqrt{1-\bar\alpha_{t_{i+1}}}\,\hat{\boldsymbol{\epsilon}},
\label{eq:ddim}
\end{equation}
starting from $\bx_{t_1}\sim\mathcal{N}(\mathbf{0},\mathbf{I})$, drawn independently of the channel and the observation, with $\bar\alpha_{t_{S+1}}=1$ at the clean end $t_{S+1}$. The output $\hat{\bR}=\bx_{t_{S+1}}$ is the residual prediction of the last step, and the estimate is $\hat{\bH}=\hat{\bH}_{\mathrm{A}}+s\hat{\bR}$. As a benchmark, a regression network of the same architecture is trained to minimize the MSE of $\bR$ directly and is evaluated once per frame. The diffusion and regression networks share the training settings of Table~\ref{tab:par}.

\section{Spacing, Posterior Sampling and the Start}
\label{sec:spc}
\subsection{Leading Versus Trailing Spacing}
\label{sec:spacing}
\begin{figure}[t]
\centering
\includegraphics[width=\columnwidth]{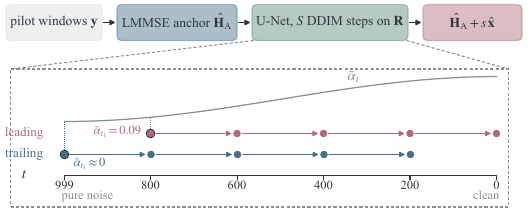}
\caption{Anchored residual diffusion estimator (top) and the timesteps of the two spacings for $S=5$ on the cosine schedule (bottom).}
\label{fig:system}
\end{figure}
The spacing rule fixes the timesteps $t_1>t_2>\dots>t_S$ on the $T=1000$ timesteps of the schedule, and the two rules below differ mainly in the first timestep, and with it in the noise level that this timestep indicates to the network. The leading rule of common DDIM implementations uses $t_i=(S-i)\lfloor T/S\rfloor$, so that $t_1=(S-1)\lfloor T/S\rfloor$, which gives $\{800,600,400,200,0\}$ for $S=5$ and $\{0\}$ for $S=1$. The trailing rule of \cite{lin2024flawed} uses $t_1=T-1$ and $t_{i+1}=t_i-\lfloor T/S\rfloor$, which gives $\{999,799,599,399,199\}$ and $\{999\}$. Under leading spacing the state part of the first network input is pure noise, but the first timestep indicates the signal power coefficient $\bar\alpha_{t_1}$, which is 0.09 for $S=5$ ($t_1=800$), 0.25 for $S=3$ ($t_1=666$) and close to one for $S=1$ ($t_1=0$). With one step it thus indicates that pure noise is an almost clean residual, and the network is queried on inputs that do not match the noise level it was trained for at $t_1$. Fig.~\ref{fig:system} illustrates both rules for $S=5$. Only the trailing rule starts where the noise level indexed by the timestep agrees with the input, at $t_1=999$ with $\bar\alpha_{t_1}\approx0$. The mismatch shrinks as $t_1$ approaches $T$, which is consistent with the accuracy of leading spacing growing with $S$ (Section~\ref{sec:steps}).

\subsection{Posterior Mean Versus Posterior Sample}
Let $\mathbf{c}$ be the conditioning, namely the anchor and the SNR, and let $\boldsymbol{\mu}=\E[\bR|\mathbf{c}]$ and $\mathcal{E}^\star=\E\|\bR-\boldsymbol{\mu}\|^2$ denote the conditional mean and the MMSE. Consider an idealized sampler with three properties. (i) $\bR$ has finite second moment. (ii) The implied prediction is exact, $\hat{\bx}_0(\bx,t)=\E[\bR\,|\,\bx_t=\bx,\mathbf{c}]$ for every $t$. (iii) The first timestep has $\bar\alpha_{t_1}=0$, and the initial state is drawn from $\mathcal{N}(\mathbf{0},\mathbf{I})$, with real and imaginary parts stacked as real channels, independently of $(\bR,\mathbf{c})$, which is then the exact terminal marginal. Our schedule has $\bar\alpha_{999}=2.4\times10^{-9}$, so for a unit-power residual the signal component of the forward state at $t=999$ lies about 86~dB below its noise component. Condition (ii) is also an idealization because our network receives the conditioning and the state as one sum, $\hat{\bH}_{\mathrm{A}}/s+\bx_t$, from which it can separate the two only approximately.
\begin{lemma}
The one-step estimate equals $\boldsymbol{\mu}$, and its MSE is $\mathcal{E}^\star$. If the reverse process is instead the exact solution of the probability flow ordinary differential equation from this endpoint to $t=0$, its output $\tilde{\bR}$ has the law $p(\bR|\mathbf{c})$, is conditionally independent of $\bR$ given $\mathbf{c}$, and $\E\|\tilde{\bR}-\bR\|^2=2\mathcal{E}^\star$.
\end{lemma}
\begin{proof}
The network is evaluated at the initial state, which is drawn independently, but by (ii) it returns the conditional mean under the forward state $\bx_{t_1}=\sqrt{\bar\alpha_{t_1}}\bR+\sqrt{1-\bar\alpha_{t_1}}\boldsymbol{\epsilon}$. At $\bar\alpha_{t_1}=0$ the forward state equals $\boldsymbol{\epsilon}$, which is independent of $(\bR,\mathbf{c})$, so $\E[\bR|\bx_{t_1}=\bx,\mathbf{c}]=\boldsymbol{\mu}$ for every $\bx$, and the single update, which returns $\hat{\bx}_0$, outputs $\boldsymbol{\mu}$ whatever the initial state. The exact probability flow maps the Gaussian initial law to $p(\bR|\mathbf{c})$ \cite{song2021ddim}. Its output is a deterministic function of $\mathbf{c}$ and of an initial state independent of $(\bR,\mathbf{c})$, so $\tilde{\bR}$ and $\bR$ are conditionally independent given $\mathbf{c}$ and share the same conditional law. Expanding both around $\boldsymbol{\mu}$, the cross term vanishes and $\E\|\tilde{\bR}-\bR\|^2=\E\|\tilde{\bR}-\boldsymbol{\mu}\|^2+\E\|\bR-\boldsymbol{\mu}\|^2=2\mathcal{E}^\star$.
\end{proof}
The lemma concerns the unnormalized MSE. Our NMSE averages the frame-normalized error $\|\hat{\bH}-\bH\|^2/\|\bH\|^2$, which weights frames by the inverse of their channel energy, so the factor of two need not hold for it exactly, and we use the lemma only to explain the direction of the effect. The factor of two is the classical excess error of a posterior sample over the conditional mean, which underlies the perception-distortion trade-off \cite{blau2018perception}. The lemma places it at the two ends of the DDIM schedule used by anchored residual estimators. In the ideal case a long deterministic trajectory therefore doubles the error of the one-step estimate, a ratio of 3.01~dB. A trained network with a finite step grid satisfies neither (ii) nor the continuous-time limit, so the measured gap between short and long trajectories combines this sampling effect with approximation errors, and the 3.01~dB value serves only as a reference.

\subsection{Step-Count Selection}
As a working hypothesis, the error at a given step count combines the initialization mismatch, which affects small $S$ under leading spacing, a sampling effect that may grow with $S$ under either spacing, and approximation errors. Section~\ref{sec:res} isolates the first of these by moving only the first timestep, while the last two are not separated. The best $S$ under leading spacing therefore depends on the SNR. Trailing spacing removes most of the mismatch, and in the ideal case one step is optimal. A trained network still depends weakly on its initial state, and averaging $K$ estimates from independent initial states reduces this variance at a fixed step count without lengthening any trajectory.

\subsection{Anchor-Error-Matched Start}
\label{sec:snrstart}
Trailing spacing starts at the pure-noise end whatever the observation. The anchor, however, already removes most of the channel, and the residual left depends on the SNR. Let $\rho(\mathrm{SNR})$ be the mean power per real element of $\bR$ at the operating SNR, measured once on 256 validation frames per SNR. It reflects the pilot SNR, the truncation of the pilot windows and the fractional-Doppler leakage that the anchor extrapolates, and falls from 1.75 at 0~dB to 0.38 at 10~dB and 0.14 at 20~dB for our $v$-prediction network. Because $s$ is set per network, it is 1.68, 0.37 and 0.13 for the noise-prediction network of Section~\ref{sec:start}. The forward state $\bx_t$ then has the residual-to-noise power ratio $q_t=\bar\alpha_t\rho/(1-\bar\alpha_t)$. For a threshold $\kappa\ge0$ the matched start is
\begin{equation}
t^\star=\min\{t:q_t\le\kappa\},\qquad \bx_{t^\star}\sim\mathcal{N}\bigl(\mathbf{0},(\bar\alpha_{t^\star}\rho+1-\bar\alpha_{t^\star})\mathbf{I}\bigr),
\label{eq:tstar}
\end{equation}
with $t^\star=T-1$ when no timestep satisfies the condition, followed by $S$ DDIM steps at $\mathrm{round}(t^\star(S-i+1)/S)$, $i=1,\dots,S$. On our schedule $\bar\alpha_{999}>0$, so $\kappa=0$ falls back to $t^\star=T-1$ and gives the trailing start. Its interior timesteps, for example $\{999,799,599,400,200\}$ for $S=5$, differ by one from the trailing grid $\{999,799,599,399,199\}$ because of rounding. The initial variance is the second moment of the forward state averaged over residuals of power $\rho$. The initial state thus matches only the total variance of the forward state, not its conditional law. It carries no information about $\bR$, while the forward state at $t^\star$ contains the component $\sqrt{\bar\alpha_{t^\star}}\bR$, whose power is $\kappa$ times that of the noise at $t^\star$. The threshold $\kappa$ therefore sets the relative size of the missing component, not a bound on the distance between the two distributions. A later start, closer to the pure-noise end, shrinks this missing component but can amplify the network error in the implied residual. A noise-prediction network forms $\hat{\bx}_0=(\bx_t-\sqrt{1-\bar\alpha_t}\,\hat{\boldsymbol{\epsilon}})/\sqrt{\bar\alpha_t}$, which multiplies its error by $1/\sqrt{\bar\alpha_t}$, about 158 at $t=999$ for the linear noise schedule of \cite{ho2020ddpm}, whereas $v$-prediction needs no such division. We choose $\kappa$ once per network on validation frames at $S=3$, from a grid with steps of about 5~dB that contains $\kappa=0$, and keep it for all SNRs and step counts. The rule needs one validation pass per network and no retraining. It differs from the SNR-matched start of \cite{fesl2024diffusion} in three ways. That estimator initializes the state with the normalized observation and matches the pilot noise. Here the state describes the residual of the anchor, so the start is matched to the anchor error power, which also contains the window truncation and the leakage, the initial state is independent noise of matched variance, and $\kappa$ is calibrated for each network.

\section{Numerical Results}
\label{sec:res}
\subsection{Setup}
\label{sec:setup}
The users of eight DeepMIMO cities (Los Angeles, Chicago, Houston, Phoenix, San Diego, Dallas, Fort Worth, Denver) are split into training (80\%), validation (10\%) and test (10\%) users, and results on the test users of these eight cities are labeled test cities. No New York user is used for training, and New York serves as the unseen city. Every frame draws a user, a speed uniformly in $[30,500]$~km/h, a heading, a uniformly random user array orientation, a base station array orientation within $\pm60^\circ$, QPSK data and noise, so the training set never repeats. The pilots of the four antennas sit at Doppler bin 32 and delay bins 40, 44, 48 and 52. The anchor statistics use 16{,}384 training frames per SNR on a grid from $-5$ to 25~dB in 5~dB steps. The anchor map has $131{,}072\times960$ complex weights per SNR point, 0.50~GB per point with half-precision real and imaginary parts, and costs 1.0~GFLOP per frame, which the latencies of Table~\ref{tab:main} include. The diffusion and regression networks share one U-Net that processes the complex channel as pairs of real channels with real convolutions and 8.26~M real parameters, and they are trained for 60{,}000 iterations with batches of 64 frames on the same SNR grid. The NMSE is $10\log_{10}$ of the frame average of $\|\hat{\bH}-\bH\|^2/\|\bH\|^2$, so the average is taken before the logarithm. For our networks it is averaged over 512 test frames per point in Section~\ref{sec:steps} and over 1024 test frames in Tables~\ref{tab:abl} to~\ref{tab:ber}, where all estimators of a table use the same frames and initial noise. The published CRDM network is evaluated on 1000 frames per SNR of its own test set. The intervals quoted in the text are 95\% confidence intervals from a paired bootstrap over frames with 2000 resamples.

\subsection{Step Count}
\label{sec:steps}
\begin{figure}[t]
\centering
\includegraphics[width=\columnwidth]{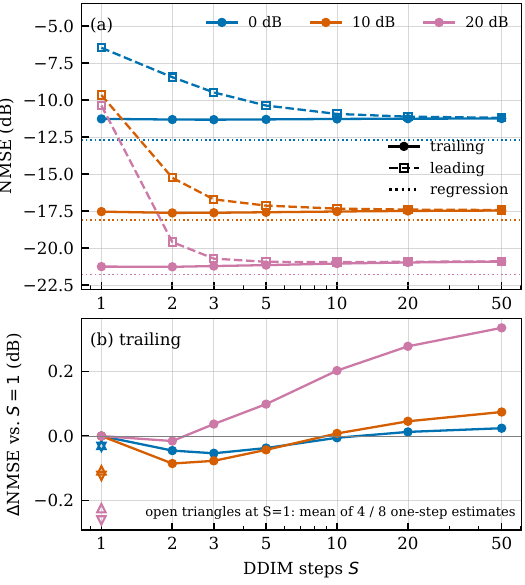}
\caption{NMSE versus the number of DDIM steps for the anchored residual diffusion estimator, test cities. (a) Leading (dashed) and trailing (solid) spacing at 0, 10 and 20~dB, with the regression network of the same architecture (dotted). (b) Trailing spacing relative to one step, and the average of 4 or 8 one-step estimates from independent initial states.}
\label{fig:steps}
\end{figure}
Fig.~\ref{fig:steps}(a) shows the NMSE against the number of DDIM steps. Under leading spacing one step fails ($-6.5$~dB at 0~dB and $-10.4$~dB at 20~dB), and the accuracy grows with $S$ up to an SNR-dependent optimum as the first timestep $t_1=(S-1)\lfloor T/S\rfloor$ approaches the pure-noise end. The best leading configuration needs 50 steps up to 15~dB and 10 steps at 20~dB, where 5 to 50 steps lie within 0.06~dB, which could suggest SNR-adaptive step counts. Under trailing spacing the curves stay within 0.4~dB of their minimum from 1 to 50 steps and lie below the leading curve at every step count. Fig.~\ref{fig:steps}(b) zooms into the trailing curves. The minimum lies at three steps for 0 and 5~dB and at two steps for 10 to 20~dB, within 0.1~dB of one step, and beyond it the error grows steadily. After 50 steps, the largest count considered, the error exceeds the one-step error by 0.07~dB at 10~dB and by 0.34~dB at 20~dB, and the two-step minimum by 0.35~dB at 20~dB. This growth, which is larger at high SNR, is consistent with the sampling effect of the lemma, but our measurements do not separate it from accumulated approximation error.

\begin{table}[t]
\centering
\caption{NMSE (dB) and Latency at 10~dB, Test Cities. Latency is per frame within a batch of 64 on one NVIDIA RTX PRO 6000 GPU shared with other jobs, so the values indicate relative cost, and it includes the anchor.}
\label{tab:main}
\footnotesize
\begin{tabular}{lcc}
\toprule
Estimator & NMSE & ms/frame \\
\midrule
LMMSE anchor & $-15.50$ & 0.29 \\
Diffusion, leading, $S=5$ (CRDM default) & $-17.12$ & 5.82 \\
Diffusion, trailing, $S=1$ & $-17.53$ & 1.45 \\
Diffusion, trailing, $S=2$ & $-17.61$ & 2.30 \\
Diffusion, trailing, $S=5$ & $-17.57$ & 5.83 \\
Diffusion, trailing, $S=20$ & $-17.48$ & 22.90 \\
Diffusion, mean of 8 trailing, $S=1$ & $-17.65$ & 9.56 \\
Diffusion, mean of 8 trailing, $S=5$ & $-17.72$ & 45.66 \\
Regression, same network & $-18.11$ & 1.45 \\
\bottomrule
\end{tabular}
\end{table}

Table~\ref{tab:main} compares accuracy and latency at 10~dB. One trailing step attains 0.41~dB lower NMSE than the five leading steps of the CRDM default and runs four times faster in measured GPU time. Averaging eight estimates from independent initial states lowers the NMSE by 0.12~dB for one-step and by 0.15~dB for five-step estimates, at eight times the network evaluations and 6.6 and 7.8 times the measured latency. At a comparable number of network evaluations, eight one-step estimates (8 evaluations) attain 0.08 to 0.47~dB lower NMSE than one ten-step trajectory (10 evaluations) at SNRs from 5 to 20~dB, and 0.03~dB lower NMSE at 0~dB. The regression network, trained directly for the MSE, attains 0.39~dB lower NMSE than the best diffusion configuration of Table~\ref{tab:main}, the mean of eight five-step trajectories, and is as fast as one diffusion step. The gap holds at every SNR from 0 to 20~dB, from 0.23 to 1.36~dB on the test cities and from 0.22 to 1.53~dB on the unseen city, and the mean of eight five-step trajectories remains the best diffusion configuration evaluated at every SNR and in both city sets. In the tested configurations the iterative sampler therefore adds no accuracy over a direct estimator of the conditional mean. On the frames of Table~\ref{tab:abl}, the regression network attains 1.40, 0.59 and 0.53~dB lower NMSE than one trailing step at 0, 10 and 20~dB, with intervals within $\pm0.10$~dB.

\subsection{Contribution of the First Timestep}
\label{sec:abl}
\begin{table*}[t]
\centering
\caption{NMSE (dB) of DDIM Spacings. Within each estimator the trained network, frames and initial noise are the same. ``First'' keeps the leading timesteps but moves the first one to 999. Anchored residual diffusion: test cities, 1024 frames. CRDM: published network on its own test set, 1000 frames.}
\label{tab:abl}
\footnotesize
\begin{tabular}{llcccccc}
\toprule
& & \multicolumn{3}{c}{Anchored residual diffusion} & \multicolumn{3}{c}{CRDM} \\
Spacing & Timesteps & 0~dB & 10~dB & 20~dB & 0~dB & 5~dB & 10~dB \\
\midrule
Leading, $S=5$ & 800, 600, \dots, 0 & $-10.55$ & $-17.10$ & $-20.96$ & $-13.76$ & $-14.69$ & $-15.72$ \\
First, $S=5$ & 999, 600, \dots, 0 & $-11.54$ & $-17.63$ & $-21.29$ & $-14.57$ & $-16.45$ & $-17.05$ \\
Trailing, $S=5$ & 999, 799, \dots, 199 & $-11.50$ & $-17.56$ & $-21.18$ & $-14.58$ & $-16.68$ & $-16.99$ \\
\midrule
Leading, $S=3$ & 666, 333, 0 & $-9.63$ & $-16.67$ & $-20.75$ & $-13.11$ & $-13.87$ & $-14.86$ \\
First, $S=3$ & 999, 333, 0 & $-11.56$ & $-17.70$ & $-21.45$ & $-14.21$ & $-16.06$ & $-16.91$ \\
Trailing, $S=3$ & 999, 666, 333 & $-11.52$ & $-17.59$ & $-21.25$ & $-14.53$ & $-16.49$ & $-16.99$ \\
\bottomrule
\end{tabular}
\end{table*}

Trailing spacing changes the first timestep and the interior timesteps at once. Table~\ref{tab:abl} separates the two by keeping the leading spacing and moving only its first timestep to 999. For our estimator this alone recovers the whole gain of trailing spacing. With five steps it attains 0.98, 0.53 and 0.33~dB lower NMSE than leading spacing at 0, 10 and 20~dB, and even 0.03 to 0.11~dB lower NMSE than full trailing spacing. With three steps it attains 0.04 to 0.21~dB lower NMSE than trailing spacing. For the published CRDM network with five steps, moving the first timestep gains 0.81~dB at 0~dB, 1.76~dB at 5~dB and 1.33~dB at 10~dB, against 0.82, 1.99 and 1.27~dB for trailing spacing. The remaining trailing timesteps add 0.23~dB only at 5~dB with five steps, and 0.08 to 0.43~dB with three steps. Three trailing steps of that network attain 0.77 to 1.79~dB lower NMSE than its published five leading steps from 0 to 10~dB. Table~\ref{tab:ci} gives the 95\% paired bootstrap intervals of these differences. For our estimator they are within $\pm0.07$~dB, and for the CRDM network they exclude zero for every first-timestep gain and for the gains of the remaining trailing timesteps at 5~dB with five steps and at 0 and 5~dB with three steps. The first timestep is therefore the primary contributor to the accuracy that leading spacing loses at small step counts. Moving it changes both the noise level indexed by the first timestep and the subsequent DDIM trajectory, so this comparison does not isolate the input mismatch from the change of trajectory, but it is consistent with the mismatch of the first step described in Section~\ref{sec:spacing}. The best trailing step count of the CRDM network remains five or ten, within 0.3~dB of three steps (Section~\ref{sec:v1}).

\begin{table*}[t]
\centering
\caption{NMSE Differences (dB) Between the Spacings of Table~\ref{tab:abl} With 95\% Paired Bootstrap Intervals Over Frames. Negative values favor the first-named spacing.}
\label{tab:ci}
\footnotesize
\setlength{\tabcolsep}{3pt}
\resizebox{\textwidth}{!}{\begin{tabular}{lcccccc}
\toprule
& \multicolumn{3}{c}{Anchored residual diffusion, test cities} & \multicolumn{3}{c}{CRDM} \\
Difference & 0~dB & 10~dB & 20~dB & 0~dB & 5~dB & 10~dB \\
\midrule
First $-$ leading, $S=5$ & $-0.98$ [$-1.02$, $-0.94$] & $-0.53$ [$-0.55$, $-0.50$] & $-0.33$ [$-0.34$, $-0.31$] & $-0.81$ [$-1.38$, $-0.29$] & $-1.76$ [$-2.31$, $-1.25$] & $-1.33$ [$-2.12$, $-0.73$] \\
Trailing $-$ first, $S=5$ & $+0.03$ [$+0.03$, $+0.04$] & $+0.07$ [$+0.07$, $+0.08$] & $+0.11$ [$+0.10$, $+0.11$] & $-0.01$ [$-0.23$, $+0.16$] & $-0.23$ [$-0.45$, $-0.05$] & $+0.06$ [$-0.21$, $+0.27$] \\
First $-$ leading, $S=3$ & $-1.92$ [$-1.99$, $-1.86$] & $-1.03$ [$-1.07$, $-0.98$] & $-0.70$ [$-0.73$, $-0.67$] & $-1.10$ [$-1.65$, $-0.51$] & $-2.18$ [$-2.77$, $-1.67$] & $-2.05$ [$-2.88$, $-1.39$] \\
Trailing $-$ first, $S=3$ & $+0.04$ [$+0.03$, $+0.04$] & $+0.11$ [$+0.11$, $+0.12$] & $+0.21$ [$+0.20$, $+0.22$] & $-0.32$ [$-0.57$, $-0.10$] & $-0.43$ [$-0.66$, $-0.22$] & $-0.08$ [$-0.38$, $+0.17$] \\
\bottomrule
\end{tabular}}
\end{table*}

\subsection{Anchor-Error-Matched Start and a Noise-Prediction Network}
\label{sec:start}
\begin{table}[t]
\centering
\caption{NMSE (dB), Test Cities, 1024 Frames With the Same Initial Noise. The noise-prediction ($\epsilon$-prediction) network is the same U-Net trained for 30{,}000 iterations with the same anchor and data. Tuned: one step from a start chosen on validation frames, for all SNRs or per SNR. Matched $S$: $S$ steps from the anchor-error-matched start. Tuned and matched starts use the initial variance of (\ref{eq:tstar}).}
\label{tab:snr}
\setlength{\tabcolsep}{2.5pt}
\footnotesize
\begin{tabular}{lcccccc}
\toprule
& \multicolumn{3}{c}{$v$-prediction, cosine} & \multicolumn{3}{c}{$\epsilon$-prediction, linear} \\
Scheme & 0~dB & 10~dB & 20~dB & 0~dB & 10~dB & 20~dB \\
\midrule
Leading 1 & $-6.55$ & $-9.64$ & $-10.41$ & $-6.49$ & $-9.51$ & $-10.25$ \\
Leading 3 & $-9.63$ & $-16.67$ & $-20.75$ & $-6.18$ & $-8.31$ & $-8.58$ \\
Leading 5 & $-10.55$ & $-17.10$ & $-20.96$ & $-0.78$ & $-0.72$ & $-0.56$ \\
First 3 & $-11.56$ & $-17.70$ & $-21.45$ & $+15.98$ & $+16.64$ & $+16.86$ \\
Trailing 1 & $-11.47$ & $-17.52$ & $-21.27$ & $+16.02$ & $+16.68$ & $+16.90$ \\
Trailing 3 & $-11.52$ & $-17.59$ & $-21.25$ & $+15.30$ & $+16.06$ & $+16.30$ \\
Tuned, all SNRs & $-11.47$ & $-17.52$ & $-21.27$ & $-8.26$ & $-15.01$ & $-18.52$ \\
Tuned, per SNR & $-11.47$ & $-17.52$ & $-21.27$ & $-8.27$ & $-15.13$ & $-19.11$ \\
Matched 1 & $-11.47$ & $-17.52$ & $-21.27$ & $-8.30$ & $-15.11$ & $-19.08$ \\
Matched 3 & $-11.52$ & $-17.59$ & $-21.25$ & $-8.26$ & $-15.06$ & $-19.07$ \\
\midrule
Start $t^\star$ & 999 & 999 & 999 & 309 & 172 & 107 \\
LMMSE anchor & $-8.50$ & $-15.43$ & $-19.53$ & $-8.50$ & $-15.43$ & $-19.53$ \\
Regression & $-12.86$ & $-18.11$ & $-21.81$ & $-12.86$ & $-18.11$ & $-21.81$ \\
\bottomrule
\end{tabular}
\end{table}

Table~\ref{tab:snr} applies the matched start to our network and to a noise-prediction network with the same U-Net, anchor and data, trained for 30{,}000 iterations with the linear noise schedule of \cite{ho2020ddpm}, whose terminal signal power is $\bar\alpha_{999}=4.0\times10^{-5}$. For the $v$-prediction network every rule selects the pure-noise end. The tuned starts are $t=999$, and the validation search for the anchor-error-matched start selects $\kappa=0$, the lower end of its grid. No timestep satisfies $q_t\le0$, so the rule falls back to $t=T-1$ and reproduces trailing spacing. One matched step is therefore identical to one trailing step, and three or five matched steps stay within 0.01~dB of trailing spacing, whose interior timesteps differ by one. A search on a grid that starts at $\kappa=10^{-3}$ also selects its lowest value, and its matched start, closer to the clean end, then has 0.004 to 0.10~dB higher NMSE than trailing spacing. The rule thus recovers trailing spacing from the observation, without improving on it.

For the noise-prediction network every tested spacing that starts at $t=999$ fails with an NMSE of $+15$ to $+17$~dB, which is the amplification $1/\sqrt{\bar\alpha_{999}}$ of Section~\ref{sec:snrstart} acting on its noise prediction. Leading spacing avoids the pure-noise end but reaches at best $-6.5$ to $-10.2$~dB, with one step at $t=0$. A grid search on validation frames selects $\kappa=1$, where the missing signal component equals the noise in power ($-14.25$~dB on validation frames, against $-13.75$~dB at $\kappa=3$ and $-13.95$~dB at $\kappa=0.1$). The optimum is flat on its lower side, since $\kappa=0.3$ is within 0.01~dB. The matched start then falls with the SNR, from 309 at 0~dB to 107 at 20~dB, as the anchor residual shrinks. With one step it attains 1.82, 5.60 and 8.83~dB lower NMSE than the best fixed spacing at 0, 10 and 20~dB on the test cities and 1.88, 5.86 and 8.89~dB on the unseen city, with intervals within $\pm0.5$~dB, and three matched steps add nothing, since their NMSE is within 0.05~dB of that of one step. The NMSE of the matched network is 0.20 to 0.45~dB higher than that of the LMMSE anchor it refines on the test cities and 0.20 to 0.39~dB higher on the unseen city, and 2.7 to 5.0~dB higher than that of the regression network. It was trained for half as many iterations as the $v$-prediction and regression networks, and part of these gaps may come from training. The noise-prediction network therefore serves only to check where the rule places the start and is not a competitive estimator.

A tuned interior start is the natural competitor. With one step and the same variance-matched initial state, a single start chosen on validation frames from a grid of 14 timesteps between 25 and 999 for all SNRs selects $t=300$, and one start chosen per SNR selects 400, 200 and 150 at 0, 10 and 20~dB. The matched start lies at 309, 172 and 107, and its NMSE comes within 0.04~dB of that of the per-SNR starts on the test cities and within 0.06~dB on the unseen city, from one parameter instead of three. Against the tuned start for all SNRs it gains 0.04, 0.09 and 0.56~dB at 0, 10 and 20~dB on the test cities (interval at 20~dB [0.52, 0.60]~dB) and 0.05, 0.06 and 0.38~dB on the unseen city. Most of its gain over the fixed spacings of Table~\ref{tab:snr} therefore comes from starting at an interior timestep, and the matched rule adds the SNR dependence of the start, taken from the anchor error power rather than tuned per SNR. At 5 and 15~dB, two SNRs not used for calibration, $\rho$ is interpolated log-linearly from its values at 0, 10 and 20~dB to 0.79 and 0.22, and with the same $\kappa$ the matched start lies at 236 and 137. The oracle start, the best on a fine grid chosen on the test frames themselves, lies at 275 and 175 on both the test cities and the unseen city, and the NMSE of the matched start is within 0.05~dB of that of the oracle start in all four cases. At 15~dB the matched start attains 0.27~dB lower NMSE on the test cities and 0.21~dB lower NMSE on the unseen city than the tuned start $t=300$, and at 5~dB its NMSE is 0.01 and 0.02~dB higher. Scaling $\rho$ per frame by the pilot-window residual of the anchor changes the NMSE by at most 0.04~dB in either direction, so the frame-level variation adds little. The threshold is not universal. $\kappa=10^{-3}$ would cost the noise-prediction network 12~dB on validation frames and $\kappa=1$ would cost the $v$-prediction network 2.6~dB, so $\kappa$ has to be chosen per network, which one validation pass does.

\subsection{Bit Error Rate}
\label{sec:ber}
The channels of our scenarios are poorly suited to four independent streams. In the two-dimensional discrete Fourier transform (2D-DFT) domain, where the DD convolution becomes a per-bin product, the $4\times4$ channel of a bin carries on average 97.0\% of its energy in the strongest singular mode and 0.0016\% in the weakest on the test cities, and the weakest mode carries more than 1\% of the energy in none of the evaluated bins of either city set. We therefore measure the BER in a single-stream configuration. All four antennas still send their pilots, so all 16 links are estimated, but only antenna~1 carries QPSK data, which a per-bin LMMSE combiner over the four receive antennas detects after the pilots have been cancelled with the estimated channel. No transmitted data is used by the receiver. The anchor, the diffusion network and the regression network are retrained for this configuration, the networks for 30{,}000 iterations, with speeds up to 800~km/h, and they are tested up to 500~km/h, the speed covered by the pilot window $k_{\max}=7$.

\begin{table}[t]
\centering
\caption{Single-Stream Configuration: Uncoded BER With the Estimated Channels and NMSE in dB, 1024 Frames per Point. Each BER rests on 15.6 million bits, with 93 (perfect CSI, test, 10~dB) to 14{,}760 (LMMSE, unseen, 10~dB) bit errors, and the zero for perfect CSI means no bit error. Lead.~5 and trail.~1 denote diffusion with five leading or one trailing step.}
\label{tab:ber}
\setlength{\tabcolsep}{3pt}
\footnotesize
\begin{tabular}{lcccc}
\toprule
& \multicolumn{2}{c}{Test cities} & \multicolumn{2}{c}{Unseen city} \\
& 10~dB & 20~dB & 10~dB & 20~dB \\
\midrule
LMMSE & $2.40{\times}10^{-4}$ & $1.68{\times}10^{-4}$ & $9.43{\times}10^{-4}$ & $1.84{\times}10^{-4}$ \\
Diff., lead.~5 & $1.24{\times}10^{-4}$ & $2.51{\times}10^{-5}$ & $5.95{\times}10^{-4}$ & $5.51{\times}10^{-5}$ \\
Diff., trail.~1 & $1.08{\times}10^{-4}$ & $2.68{\times}10^{-5}$ & $5.90{\times}10^{-4}$ & $3.48{\times}10^{-5}$ \\
Regression & $9.18{\times}10^{-5}$ & $5.84{\times}10^{-5}$ & $5.28{\times}10^{-4}$ & $3.30{\times}10^{-5}$ \\
Perfect CSI & $5.94{\times}10^{-6}$ & $0$ & $2.80{\times}10^{-4}$ & $0$ \\
\midrule
NMSE, LMMSE & $-13.83$ & $-16.19$ & $-14.07$ & $-16.48$ \\
NMSE, lead.~5 & $-15.78$ & $-18.37$ & $-16.22$ & $-18.71$ \\
NMSE, trail.~1 & $-16.18$ & $-18.80$ & $-16.60$ & $-19.06$ \\
NMSE, regr. & $-16.59$ & $-19.34$ & $-17.07$ & $-19.69$ \\
\bottomrule
\end{tabular}
\end{table}

Table~\ref{tab:ber} shows that the NMSE ordering of the step study carries over to this configuration. The observed BER ordering agrees with it in three of the four reported conditions and is reversed at 20~dB on the test cities. One trailing step lowers the uncoded BER against the five leading steps by 12.8\% at 10~dB on the test cities and by 36.8\% at 20~dB on the unseen city, changes it by less than 1\% at 10~dB on the unseen city and raises it by 6.9\% at 20~dB on the test cities. In the three conditions where the orderings agree, the regression network lowers the BER of one trailing step by a further 5.1 to 14.7\%. At 20~dB on the test cities the learned estimators produce only 393 to 914 bit errors, and the BER of the regression network is 2.2 times that of one trailing step. These are point estimates without intervals, and we do not establish the statistical significance of the BER differences.

\subsection{The Published CRDM Network}
\label{sec:v1}
We also evaluated the published CRDM network \cite{zheng2026crdm} on its own test set, an OTFS model with $4\times4$ antennas and $64\times128$ bins in which all paths of a link share one Doppler shift, changing only the timestep spacing of the DDIM scheduler. Table~\ref{tab:v1} reports the NMSE over 1000 frames per SNR, computed with the same aggregation as above. Table~\ref{tab:abl} uses the same frames with an explicit DDIM loop and other initial-noise draws, and its trailing five-step values differ from Table~\ref{tab:v1} by up to 0.22~dB at 5~dB, which reflects how much this network depends on its initial state. Trailing spacing is at least as accurate as leading spacing for every step count at every SNR. In Table~\ref{tab:v1}, three trailing steps attain 0.81~dB lower NMSE at 0~dB, 1.54~dB at 5~dB and 1.33~dB at 10~dB than the published five-step leading configuration. The two networks use different channel models, so we compare only their gains, not their absolute NMSE. A single trailing step of this network has 2.3 to 3.1~dB higher NMSE than its best trailing step count, against at most 0.1~dB for our estimator on the test cities, and its trailing accuracy saturates from three steps, with the best value at five or ten steps and differences within 0.3~dB beyond three steps. A first evaluation with 250 frames per SNR showed one reversal, leading spacing ahead by 1.0~dB at 8~dB with ten steps, which disappears with 1000 frames.

\begin{table}[t]
\centering
\caption{Published CRDM Network on Its Original Test Set, NMSE (dB).}
\label{tab:v1}
\setlength{\tabcolsep}{2.5pt}
\footnotesize
\begin{tabular}{lccccccc}
\toprule
& \multicolumn{3}{c}{Leading} & \multicolumn{4}{c}{Trailing} \\
SNR & $S=3$ & $S=5$ & $S=10$ & $S=1$ & $S=3$ & $S=5$ & $S=10$ \\
\midrule
0~dB & $-13.11$ & $-13.76$ & $-14.47$ & $-12.30$ & $-14.57$ & $-14.61$ & $-14.47$ \\
3~dB & $-13.93$ & $-14.59$ & $-14.97$ & $-13.05$ & $-15.55$ & $-15.59$ & $-15.52$ \\
5~dB & $-13.93$ & $-14.75$ & $-15.52$ & $-13.46$ & $-16.29$ & $-16.46$ & $-16.55$ \\
8~dB & $-14.05$ & $-14.81$ & $-15.57$ & $-13.51$ & $-15.88$ & $-16.00$ & $-16.02$ \\
10~dB & $-14.86$ & $-15.67$ & $-16.26$ & $-14.44$ & $-17.00$ & $-17.06$ & $-16.96$ \\
\bottomrule
\end{tabular}
\end{table}

\section{Conclusion}
\label{sec:con}
For our estimator, the accuracy that seemed to require more DDIM steps is recovered by starting the reverse process at the pure-noise end, and moving only the first timestep suffices. The published CRDM network gains from the same move of the first timestep, although its best trailing step count remains five or ten. With trailing spacing one to three steps are best for our estimator, and longer trajectories raise the error. The rise is consistent with a move from the posterior mean toward a posterior sample, so part of the appeal of SNR-adaptive step counts may be compensation for the spacing error, although our experiments do not separate the two effects. Choosing the start from the error power of the LMMSE anchor recovers trailing spacing for a $v$-prediction network and, for a noise-prediction network on which every start at the pure-noise end fails, matches a per-SNR tuned start with one parameter and predicts how the best start moves with the SNR, although that network is less accurate than its anchor. When the MSE is the target, averaging a few short estimates is a better use of computation than one long trajectory, and a regression network of the same size is more accurate than every diffusion configuration we report from 30 to 500~km/h.

\bibliographystyle{IEEEtran}
\bibliography{refs}
\end{document}